# Binding of transcription factors adapts to resolve information-energy tradeoff

Yonatan Savir[1,*], Jacob Kagan[2] and Tsvi Tlusty[3,*]

[1]*Department of Systems Biology, Harvard Medical School, Boston, 02115, USA*
[2]*Department of Mathematics, Weizmann Institute of Science, Rehovot 76100, Israel*
[3]*Simons Center for Systems Biology, Institute for Advanced Study, Princeton, New Jersey 08540, USA*
[*]*Correspondence: tlusty@ias.edu, yonatan_savir@hms.harvard.edu*

**Abstract**

We examine the binding of transcription factors to DNA in terms of an information transfer problem. The input of the noisy channel is the biophysical signal of a factor bound to a DNA site, and the output is a distribution of probable DNA sequences at this site. This task involves an inherent tradeoff between the information gain and the energetics of the binding interaction – high binding energies provide higher information gain but hinder the dynamics of the system as factors are bound too tightly. We show that adaptation of the binding interaction towards increasing information transfer under a general energy constraint implies that the information gain per specific binding energy at each base-pair is maximized. We analyze hundreds of prokaryote and eukaryote transcription factors from various organisms to evaluate the discrimination energies. We find that, in accordance with our theoretical argument, binding energies nearly maximize the information gain per energy. This work suggests the adaptation of information gain as a generic design principle of molecular recognition systems.

## 1 Introduction

The binding of transcription factors (TFs) to specific DNA sites is essential to the regulation of gene expression [1]. TFs have to cope with the task of recognizing a specific DNA subsequence within long strands, often containing numerous similar subsequences. The main determinant of TF specificity is the sequence-specific DNA binding energy, termed *discrimination energy* [2, 3]. The proper function of the transcription regulation network requires that the TFs accommodate two conflicting needs: on the one hand, specificity favors high discrimination energies, on the other hand, search time and dynamical flexibility of the network favors low binding energies [4-6].

TF specificity is an extensively studied field [7]. Many studies take a biophysical approach by relating the occurrence of a TF on a DNA site to the corresponding binding free energy [2, 3, 8-15]. To estimate the absolute binding probability, one has to also take into account the TF concentration (the chemical potential) [8]. However, this concentration dependence cancels out in *relative* binding probabilities considered in the present work, where it suffices to evaluate solely the discrimination energy [7]. The discrimination energy is the free energy difference between the main target motif and other, similar DNA motifs. It can be estimated from binding motif statistics under the assumption of an additive discrimination energy, where the contribution of each base pair (bp) is summed independently [2, 16]. The resulting discrimination energy per bp, $\varepsilon$, is in the range of $\varepsilon \approx$ 2-4 $k_B T$. [17, 18], which is consistent with experimental results [19].

The particular value of the discrimination energy, $\varepsilon$, raises the question as to whether it is the outcome of biophysical constraints or perhaps of adaptation to improve the performance of the transcription regulation network. Values of $\varepsilon \approx 2$ $k_B T$ per bp, in a 15-bp long TF binding site were suggested to achieve both specific and fast recognition [4, 5]. However, besides the search time, there could be other factors which counterbalance the specificity. Recent studies suggest that the dynamics of signaling molecules play a role in regulation [6, 20]. For example, the dynamics of the yeast TF Msn2 can vary between transient increase to pulses, leading to different responses [21]. These types of dynamic control could be hindered if the binding energies are too high. The existence of multiple tradeoffs raises the question of whether the values of discrimination energies are the result of some universal tradeoff or that adaptation is specific to the each case by itself.

In this work, we take an alternative information theory approach that is not sensitive to the underlying dynamic control considerations (such as avoiding long search time). We formulate the problem as an optimization problem in which the information regarding the sequence is maximized under a general energy constraint. Then, per given value of overall energy, we derive a simple expression for the discrimination energy that maximizes the information obtained by a binding event. We show that this amounts to maximizing the information gain per discrimination energy for each bp. To test this argument, we evaluate the discrimination energies and target length from the binding motif distribution of hundreds of TFs from prokaryotes and eukaryotes. The analyzed data suggest that the discrimination energies are distributed around this optimal value.

## 2 Optimal information gain under energy constraint

Upon binding to a DNA site, a TF conveys information regarding the sequence identity of that site. In other words, it reduces the uncertainty regarding the bound sequence by sharpening the probability distribution of possible DNA sequences (Fig. 1A). In the following, we quantify the information regarding this DNA sequence gained by the knowledge that it is bound by a TF. In a randomly distributed genome, a TF that recognizes a site of length $N$ bp can in principle bind to any of the $4^N$ possible target sequences. Each possible sequence is specified by a vector $\mathbf{s} = \{s_k\}$, $k = 1,..., N$, where $s_k$ is the base at position $k$.

We consider a simple two-state model, where at each position $k$ there is a favored 'consensus' base $c_k$, and any of the three possible mismatches between the bp $s_k$ and its consensus $c_k$, incurs an energy cost $\varepsilon_k$ [2, 5]. The overall binding energy is therefore linear,

$$E(\mathbf{s}) = \sum_{k=1}^{N} \varepsilon_k \delta(s_k, c_k), \quad (1)$$

where the zero energy is set at an all-mismatch state and a perfect match yields a discrimination energy $E(\mathbf{s} = \mathbf{c}) = \sum_k \varepsilon_k$ (in units of $k_B T$). In a random genome, the probability that a bound TF is bound to a sequence $\mathbf{s} = \{s_k\}$, $p(\mathbf{s}) \equiv P(\mathbf{s} \mid \text{TF bound})$ is given by the Boltzmann factor, which decomposes into a product of position probabilities,

$$p(\mathbf{s}) = \frac{e^{E(\mathbf{s})}}{Z} = \prod_{k=1}^{N} p(s_k), \quad p(s_k) = \frac{e^{\varepsilon_k \delta(s_k, c_k)}}{e^{\varepsilon_k} + 3}, \quad (2)$$

where the partition function is $Z = \sum_{\mathbf{s}} e^{E(\mathbf{s})} = \prod_k \left(e^{\varepsilon_k} + 3\right)$.

Prior to binding, the entropy of a random $N$-bp sequence is $H_0 = 2N$ bits, whereas after binding it is reduced to $H_b = -\sum_{\mathbf{s}} p(\mathbf{s}) \log_2 p(\mathbf{s})$ bits. The overall information regarding the DNA sequence gained by observing a bound TF is therefore

$$I_N = H_0 - H_b = 2N + \sum_{\mathbf{s}} p(\mathbf{s}) \log_2 p(\mathbf{s}) \quad (3)$$

By substitution of the probability (2) in the information (3), we find that $I_N$ (in bits) is a sum over contributions from each bp,

$$I_N(\{\varepsilon_k\}) = 2N + \sum_{\mathbf{s}} \prod_{h=1}^{N} p(s_h) \log_2 \prod_{k=1}^{N} p(s_k) =$$
$$= 2N + \sum_{k=1}^{N} \sum_{\mathbf{s}} \prod_{h=1}^{N} p(s_h) \log_2 p(s_k) = \sum_{k=1}^{N} \left[ 2 + \sum_{s_k} p(s_k) \log_2 p(s_k) \right] = \sum_{k=1}^{N} I(\varepsilon_k), \quad (4)$$

where the bp contributions are given by,

$$I(\varepsilon_k) = 2 + \sum_{s_k} p(s_k) \log_2 p(s_k) = \frac{1}{\ln 2} \left( \ln \frac{4}{3 + e^{\varepsilon_k}} + \frac{\varepsilon_k}{1 + 3e^{-\varepsilon_k}} \right). \quad (5)$$

As expected, the information vanishes at the limit of small discrimination energies, $\varepsilon_k = 0$, since no sequence is favored. At the other extreme of large discrimination energies, $\varepsilon_k \to \infty$, the TF is most likely to bind to the consensus sequence with gained information of $2N$ bits. The information curve $I(\varepsilon)$ (5) takes a sigmoidal shape saturating around $\varepsilon \approx 7$ $k_BT$ (Fig. 1B).

Transcription factors (TFs) operate in a complex biophysical environment with many opposing demands. The discrimination energy has to be large enough to discern between numerous sequences, as well as to prevent spontaneous detachment from the DNA, and yet, it has to be also small enough to allow dynamic flexibility. Proteins with high binding energy will tend to get trapped along the DNA rendering feedback and control inefficient. The length $N$ of the biding site may also vary and is likely to coevolve with the overall energy $E$. Such energetic constraints may vary among different species resulting in diverse values of $E$ [17].

Assuming that the total discrimination energy $E$ is governed by external constraints, such as TF concentrations, genome length, regulation network topology and DNA packing, we consider the following optimization problem: Given a constraint over $E$, what should be the discrimination energy profile $\varepsilon_k$, and length $N$ that yield maximal information $I_N$? – The optimal profile is found by maximizing the Lagrangian, $L = I_N - \lambda E = \sum_k I(\varepsilon_k) - \lambda \sum_k \varepsilon_k$, with respect to $\varepsilon_k$. The solution is given by the condition $dI(\varepsilon_k)/d\varepsilon_k = \lambda$. The optimal profile is therefore $\varepsilon_k = \varepsilon^*$, and to find its length $N$, we maximize $I_N = N \cdot I(E/N)$ with respect to $N$ (Fig. 1B, 1C). *The optimal value $\varepsilon^*$ maximizes the information gain per energy of each bp, $I/\varepsilon$ (Fig 1D),*

$$\frac{dI_N}{dN} = \frac{d}{dN}\left(N \cdot I\left(\frac{E}{N}\right)\right) = 0 \quad \Leftrightarrow \quad \frac{d}{d\varepsilon}\left(\frac{I}{\varepsilon}\right) = 0. \tag{6}$$

The solution of (6) is

$$\varepsilon_k = \varepsilon^* \approx 3.35 \, k_BT, \quad N^* = E/\varepsilon^*. \tag{7}$$

The optimal length $N^*$ therefore scales linearly with the energy $E$, and the maximal information gain is $I(\varepsilon^*)/\varepsilon^* \approx 0.42$ bits/$k_BT$. At the optimal configuration, the sequence probability (2) scales exponentially in the number of mismatches $m = N - \sum_k \delta(s_k, c_k)$,

$$p(\mathbf{s}) = \frac{e^{-m\varepsilon^*}}{\left(1 + 3e^{-\varepsilon^*}\right)^{E/\varepsilon^*}}. \tag{8}$$

A useful measure for the optimality of information transfer is the information $I_N$ per given energy $E$ normalized by its maximum,

$$F(\varepsilon) = \frac{I_N(E)}{I_{N^*}(E)} = \frac{I(\varepsilon)/\varepsilon}{I(\varepsilon^*)/\varepsilon^*}. \tag{9}$$

Thus, the measure to be optimized has a *twofold* interpretation: maximizing the information under energy constraint is equivalent to optimizing the information per bp per energy (Fig 1D). The simple optimization argument predicts that the measured distributions of discrimination energies will be centered around the peak at $F(\varepsilon^*) = 1$. The information gain is close to maximal, $F(\varepsilon) \geq 90\%$, in the range $\varepsilon = 2.4 - 4.8\ k_BT$ peaking at $3.35\ k_BT$. Such a range enables tuning of each bp to accommodate specific biophysical requirements. In the following, this optimality prediction is examined by evaluating the discrimination energy from TF binding statistics.

We note that the simplified two-state model we apply disregards possible interactions among the positions of the binding site (linear binding energy, (1)-(2)), which are known to be relevant in inferring the binding of specific TFs [22, 23]. We use this approximation since our aim here is to elucidate the basic information gain mechanism. Nevertheless, our information gain argument can be extended to the general case of non-linear binding energy with corresponding analysis of the binding motifs.

### 3 Inferring discrimination energies from motif statistics

The outcome of an ideal TF binding assay is a list of $S$ sequences where the TF binds. The binding motif is characterized by the frequency matrix, whose entries $n_{kb}$ are the occurrence of nucleotide $b$ at position $k$. A non-uniform $k$-th row of the matrix indicates a tendency of the TF to bind specific nucleotides more than others, namely a non-zero discrimination energy $\varepsilon_k$. Measurements of direct TF binding such as protein binding microarrays (PBMs) [24] and systematic evolution of ligands by exponential enrichment (SELEX) [25] allow one to infer the biophysical binding parameters [9, 12]. An indirect approach to estimate bp frequency of a binding motif is via bioinformatics methods, which basically align known binding sites with ones from other genomes. While this method is useful in predicting binding sequences, the conservation of binding sites is not directly related to binding energies and mixes biophysical properties with evolutionary dynamics [14, 15].

A non-uniform distribution of binding sites might result from sampling errors, and not from actual discrimination energy. To account for this inherent sampling noise, one has to evaluate the null distribution that yields an observed energy value (Fig 2A, 2B). Figure 2B illustrates how some positions have frequencies that fall within what is expected from a random model. To filter out the effect of noise, we use the KL null distribution, which is in our case the one of evenly distributed base-pairs for a given sample size (Fig 2B, 2C). The average value of the null distribution can be easily calculated without simulations. Thus, for the sake of reproducibility, we use it a basis for our cutoff (Fig 2C). A binding position is considered *significant* only if its $D_{KL}$ exceeds that of $S$ samples drawn from a random nucleotide distribution [26]. $N$ denotes the number of significant positions, the effective length of the binding site.

Once the significant positions are determined, the discrimination energies are evaluated. The standard method infers the binding probability $p_{kb} \equiv p(s_k = b)$ from the $n_{kb}$ matrix using uniform distribution as a Bayesian prior [2], which yields a Dirichlet distribution,

$$\text{Prob}(p_{kb} = x_{kb}) = \text{Di}(\{n_{kb}+1\},\{x_{kb}\}) \equiv \prod_{b=A,C,G,T} x_{kb}^{n_{kb}}. \tag{10}$$

The resulting expected binding probabilities are $Ep_{kb} = (n_{kb} + 1)/\sum_c (n_{kc} + 1) = (n_{kb} + 1)/(S + 4)$, which amounts to adding a 'pseudo count' to all entries of the occurrence matrix, $n_{kb} \to n_{kb} + 1$. While a common estimator of the discrimination energies is the logarithm of the expected probability, $\varepsilon_{kb} = \ln Ep_{kb}$ [2, 3, 7, 17], a better estimator is the expected logarithm of the probability,

$$\varepsilon_{kb} = E \ln p_{kb} = \psi(n_{kb}+1) - \psi(S+4), \tag{11}$$

where $\psi$ is the digamma function. At each position $k$, the average discrimination energy $\varepsilon_k$ is estimated by the average difference between the maximal $\varepsilon_{kb}$ and the three other values,

$$\varepsilon_k = \frac{1}{3}\sum_c \left(\max(\varepsilon_{kb} - \varepsilon_{kc})\right). \tag{12}$$

The total discrimination energy, $E$, is given by summing over all significant positions, $E = \sum_k \varepsilon_k$.

It is noteworthy that the sample size $S$ may significantly bias the data. The maximal discrimination energy is obtained when all sequences exhibit the same consensus, $n_{kb} = S$. The resulting estimator for the discrimination energy (11) is bounded by the sample size,

$$\varepsilon_{\max}(S) = \psi(S+1) - \psi(1) = H(S), \tag{13}$$

where $H(S)$ is the harmonic number. Similarly, the log-likelihood bound is $\varepsilon_{\max}(S) = \ln(S + 1)$. Figures 2E and 2F demonstrate how analyzing databases with small sample size may bias the energy estimates. For example, the data from Harbison et al. [27] exhibit energies in the vicinity of the sampling bound, thereby strongly biasing the distribution (Fig 2E). In contrast, the RegulonDB database [28] shows a general trend independent of sampling size (Fig 2F). Hence, to eliminate any potential bias due to small samples, we employ a conservative approach: we first remove all the energies whose error bars exceed the bound $\varepsilon_{\max}(S)$. Then, we set a cutoff to eliminate data points with low sampling.

**4 Specific binding energies are nearly optimal**

In this study, we analyzed studies of direct TF binding measurements (Fig. 3) and databases that are based on bioinformatics (Fig.4). Figure 3 shows the results from protein binding microarrays (PBMs) [29-34] and a HT-SELEX human TF study, [30]. Figure 4 shows an analysis of TF specificities databases, a eukaryotic database, [35], and two bacterial prokaryote databases [28, 36]. Taking into account the sampling limitations, we inferred the discrimination energies $\varepsilon_k$ for a few hundred TFs from several databases in different organisms. Figure 5 shows that the results of Figures 3 and 4 are practically insensitive to the cutoff choice for significant positions. Moreover, it demonstrates that, beyond the average energy, the whole distribution of energies of a single protein also tends to fall within the optimal regime.

We find that the discrimination energies are distributed around the optimal value $\varepsilon^* \approx 3.35$ $k_BT$ expected from maximizing the information gain $I/\varepsilon$ (7). In accord, we find that the overall discrimination energy $E$ scales linearly with target length $N$. In other words, the discrimination energy per bp, $\varepsilon = E/N$, is roughly the same for all TFs of a given species regardless of their target length (Fig. 3, Fig 5). The $\varepsilon$-distributions are well within the optimal regime where the information gain is more the 80% of its maximal value, $I(\varepsilon^*)/\varepsilon^* \approx 0.42$ bits/$k_BT$. The data reveals a difference between prokaryotes and eukaryotes: the discrimination energies of prokaryotes tend to be lower than those of eukaryotes. These results place the prokaryotes on the less specific side of the tradeoff, whereas the eukaryotes tend to gain more information at an energetic cost.

## 5 Discussion

Transcription factors (TFs), as well as other DNA-binding proteins, operate in a complex environment under many conflicting constraints, such as rate vs. specificity or accuracy vs. efficiency [37-40]. In general, the discrimination energy $E$ has to be large enough to discern between various sequences and to avoid early unbinding from the DNA, and yet small enough to prevent jamming of binding sites. These constraints depend on many factors, such as TF concentration, genome size and packing, cell state, the architecture of the regulation network, TF cooperativity etc.

Here, we presented an information theory approach, which reconciles these conflicting tasks by assuming hierarchal adaptation: The overall discrimination energy $E$ is determined by the performance constraints mentioned above. Once $E$ is determined, it is partitioned among the DNA bps such that the overall information gained $I_N$ is maximal. We showed that this is equivalent to maximizing the local information gain per discrimination energy $I/\varepsilon$ by setting the local discrimination energy $\varepsilon$ around an optimal value $\varepsilon^* \approx 3.35$ $k_BT$. This simple argument is consistent with detailed analysis of hundreds of TFs from prokaryotes and eukaryotes.

The observation of a variance in the distribution of $\varepsilon$ values around $\sim 3$ $k_BT$, suggests that they are not the outcome of physical limitations but rather of adaptation. Given that the total discrimination energy is constrained, we show that in order to maximize the information $I_N$, all species should have similar discrimination energy per bp. This of course is a coarse grained approximation, which disregards many relevant details of the transcription regulation system. For example, 'global' TFs are designed to bind to many different DNA sites, and TF specificity may be enhanced by cooperativity. There are also error correction and proofreading mechanisms that compensate for potential binding errors. On the other hand, high binding energy could have negative outcomes. For example, high binding energies generally require larger binding molecules and larger DNA sites, which are both costly to produce and maintain. As a result, we find that the information gain per energy $I/\varepsilon$ is a relevant measure for the TF performance.

Previous estimates of the discrimination energies in the range of $\varepsilon \approx$ 2-3 $k_BT$ per bp [17] are consistent with our analysis. However, the present study differs in two major points: we have not used motifs whose energies are severely biased due to small sample size, and we have filtered out insignificant positions in the binding sequence which exhibit random bp statistics. While previous work discussed the tradeoff between specificity and search time [5], in the present study we discuss a general scenario in which the overall binding energy is constrained, without specifying the concrete mechanism for this limitation, such as increased search time. Our work thus provides a design principle for understating the shaping of discrimination energy profiles of other sequence-binding proteins in nature and in artificial settings.

**Figure Legends**

**Figure 1. Information gain under energy constraints.** (A) Once a transcription factor is bound to the DNA, it reduces the uncertainty regarding the sequence that it is bound to. The information per bp on the bound sequence $I$, given that the protein is bound, depends on the discrimination energy per bp, $\varepsilon$ (Eq. (5)). When $\varepsilon = 0$, binding is random and the information vanishes. At the other extreme of high $\varepsilon$, $I$ approaches 2 bits, as there are 4 possible bases. (B) The total information $I_N$ as a function of the total discrimination energy $E$ and the number of sequence positions $N$ is color coded. Since high energy could have negative consequences, we consider the optimal information gain under a given energy constraint (dashes lines). The sequence length that provides maximal information $I_N$ per given energy $N^*(E)$ is denoted by the black solid line. (C) Cross sections of constant energy from (B) as a function of $\varepsilon = E/N$ exhibit a maximum at the same value $\varepsilon^* \approx 3.35\ k_BT$. (D) The normalized cross sections of information gain at constant energy collapse onto the same curve, which is the normalized information gain per bp per energy $I/\varepsilon$.

**Figure 2 Inferring discrimination energies.** (A) The Kullback-Leibler (KL) distance of the human factor *ELK1* nucleotide distribution relative to a random distribution. The discrimination energy $\varepsilon$ – the energy loss due to deviation from the most frequent nucleotide – is evaluated from the frequency of nucleotides in the different positions (Eq. (12)). (B) In some positions, non-zero energy may be an artifact of sampling error and not of actual binding. The red histogram is the distribution of the KL-distance, $D_{KL}$, of equally distributed bases given the sample size, while the black histogram is the $D_{KL}$ distribution of the protein. Some positions fall within the noise distribution. (C) The dependence of the $D_{KL}$ values of 50% (blue circles), 75% (orange circles) and 95% (green circles) percentile of the noise distribution as a function of sample size. The red line is the twice the average of the noise distribution (the red horizontal line on 2A). We consider only *significant* positions with $D_{KL}$ above twice the average $D_{KL}$ of a binding motif with random distribution and the same sample size (red horizontal line). In the case of *ELK1*, the first three positions are insignificant. (D) The dependence of the average $\varepsilon$ on the cutoff for significant positions. (E) Distribution of discrimination energies $\varepsilon$ (left) and their dependence on the number of samples $S$ (right), for a data from Harbison et al.[27]. In this dataset, all the TFs have $S = 20$ samples and maximal observable value is $\varepsilon_{max}(S) \approx 3.6\ k_BT$ (left, red vertical line). The distribution is truncated at $\varepsilon_{max}(S)$, and is therefore biased towards lower $\varepsilon$ values. (F) Analysis of a dataset from the ReguolonDB database [41] samples. The red line denotes the maximal inferred value of the $\varepsilon_{max}(S)$ (13). To avoid bias due to low sampling that results in truncation of the discrimination energy, we consider only TFs with large enough samples (red points).

**Figure 3: The distribution of average discrimination energy per base-pair based on direct TF measurements.** Each panel depicts the distribution of discrimination energies, $\varepsilon$ (left, blue) overlaid with the normalized information per energy per base-pair, $F(\varepsilon)$. The red circles denote the range of standard deviation of the $\varepsilon$ distribution. The right panel shows the total discrimination energy $E$ and the sequence length $N$ overlaid on the color coded contour plot of $F(\varepsilon)$. (A) Human HT-SELEX measurements from Jolma et al., $S > 100$ [30] (B) PBM measurements for mouse TFs from Badis et al. [32]. (C) PBM measurements for mouse TFs from Berger et al. [34] (D) PBM measurements for yeast TFs from Zhu et al [42]. For the PBM data we have used $S = 28$.

**Figure 4: The distribution of average discrimination energy per base-pair from computational databases.** Each panel depicts the distribution of discrimination energies, $\varepsilon$ (left, blue) overlaid with the normalized information per energy per base-pair, $F(\varepsilon)$. The red circles denote the range of standard deviation of the $\varepsilon$ distribution. The right panel shows the total discrimination energy, $E$ and the sequence length $N$ overlaid on the color coded contour plot of $F(\varepsilon)$. (A)–(C) TFs from the JASPAR database, $S > 100$ [35] (D) TFs from the RegTransBase database, $S > 10$ [36] (E) TFs from the RegulonDB database, $S > 10$ [41].

**Figure 5: Variation of base-pair energy and the dependence of the average $\varepsilon$ on the significant position cutoff.** Each panel relates to data presented in Figures 3 and 4, as denoted by its title. In the left hand side, the proteins are sorted according to their average $\varepsilon$ (black line). For each protein, each base-pair may have a different value of $\varepsilon$. The green shaded area covers the values of the distribution which are within the 25%-75% percentile (as in standard box-plot). The right hand side shows the

dependence of the average $\varepsilon$ values on the cutoff for significant positions (black line). The green shaded area covers the protein distribution around the average. (Note that on the left hand side the average is over base-pairs per protein, whereas on the right hand side, the average is over proteins per cutoff) (A) Human HT-SELEX measurements from Jolma et al., $S >100$ [30] (B) PBM measurements for mouse TFs from Badis et al. [32]. (C) PBM measurements for mouse TFs from Berger et al. [34] (D) PBM measurements for yeast TFs from Zhu et al. [42] (E)–(G) TFs from the JASPAR database, $S >100$ [35] (H) TFs from the RegTransBase database, $S >10$ [36] (I) TFs from the RegulonDB database, $S >10$ [41].

**References**

1. Spitz, F. and E.E. Furlong, *Transcription factors: from enhancer binding to developmental control.* Nature Reviews Genetics, 2012. **13**(9): p. 613-626.
2. Berg, O.G. and P.H. von Hippel, *Selection of DNA binding sites by regulatory proteins: Statistical-mechanical theory and application to operators and promoters.* Journal of molecular biology, 1987. **193**(4): p. 723-743.
3. Stormo, G.D. and D.S. Fields, *Specificity, free energy and information content in protein–DNA interactions.* Trends in biochemical sciences, 1998. **23**(3): p. 109-113.
4. Lassig, M., *From biophysics to evolutionary genetics: statistical aspects of gene regulation.* BMC Bioinformatics, 2007. **8 Suppl 6**: p. S7.
5. Gerland, U., J.D. Moroz, and T. Hwa, *Physical constraints and functional characteristics of transcription factor–DNA interaction.* Proceedings of the National Academy of Sciences, 2002. **99**(19): p. 12015-12020.
6. Purvis, J.E. and G. Lahav, *Encoding and decoding cellular information through signaling dynamics.* Cell, 2013. **152**(5): p. 945-956.
7. Stormo, G.D., *Modeling the specificity of protein-DNA interactions.* Quantitative Biology, 2013. **1**(2): p. 115-130.
8. Djordjevic, M., A.M. Sengupta, and B.I. Shraiman, *A biophysical approach to transcription factor binding site discovery.* Genome research, 2003. **13**(11): p. 2381-2390.
9. Foat, B.C., A.V. Morozov, and H.J. Bussemaker, *Statistical mechanical modeling of genome-wide transcription factor occupancy data by MatrixREDUCE.* Bioinformatics, 2006. **22**(14): p. e141-e149.
10. von Hippel, P.H., *On the molecular bases of the specificity of interaction of transcriptional proteins with genome DNA*, in *Biological regulation and development*. 1979, Springer. p. 279-347.
11. Von Hippel, P.H. and O.G. Berg, *On the specificity of DNA-protein interactions.* Proceedings of the National Academy of Sciences, 1986. **83**(6): p. 1608-1612.
12. Kinney, J.B., G. Tkačik, and C.G. Callan, *Precise physical models of protein–DNA interaction from high-throughput data.* Proceedings of the National Academy of Sciences, 2007. **104**(2): p. 501-506.
13. Kinney, J.B., et al., *Using deep sequencing to characterize the biophysical mechanism of a transcriptional regulatory sequence.* Proceedings of the National Academy of Sciences, 2010. **107**(20): p. 9158-9163.
14. Mustonen, V. and M. Lässig, *Evolutionary population genetics of promoters: predicting binding sites and functional phylogenies.* Proceedings of the National Academy of Sciences of the United States of America, 2005. **102**(44): p. 15936-15941.
15. Mustonen, V., et al., *Energy-dependent fitness: a quantitative model for the evolution of yeast transcription factor binding sites.* Proceedings of the National Academy of Sciences, 2008. **105**(34): p. 12376-12381.

16. Benos, P.V., M.L. Bulyk, and G.D. Stormo, *Additivity in protein-DNA interactions: how good an approximation is it?* Nucleic Acids Res, 2002. **30**(20): p. 4442-51.
17. Wunderlich, Z. and L.A. Mirny, *Different gene regulation strategies revealed by analysis of binding motifs.* Trends Genet, 2009. **25**(10): p. 434-40.
18. Zhao, Y., D. Granas, and G.D. Stormo, *Inferring binding energies from selected binding sites.* PLoS computational biology, 2009. **5**(12): p. e1000590.
19. Maerkl, S.J. and S.R. Quake, *A systems approach to measuring the binding energy landscapes of transcription factors.* Science, 2007. **315**(5809): p. 233-7.
20. Behar, M. and A. Hoffmann, *Understanding the temporal codes of intra-cellular signals.* Current opinion in genetics & development, 2010. **20**(6): p. 684-693.
21. Hao, N. and E.K. O'Shea, *Signal-dependent dynamics of transcription factor translocation controls gene expression.* Nature structural & molecular biology, 2012. **19**(1): p. 31-39.
22. Santolini, M., T. Mora, and V. Hakim, *A general pairwise interaction model provides an accurate description of in vivo transcription factor binding sites.* PloS one, 2014. **9**(6): p. e99015.
23. Zhao, Y., et al., *Improved models for transcription factor binding site identification using nonindependent interactions.* Genetics, 2012. **191**(3): p. 781-790.
24. Mukherjee, S., et al., *Rapid analysis of the DNA-binding specificities of transcription factors with DNA microarrays.* Nature genetics, 2004. **36**(12): p. 1331-1339.
25. Tuerk, C. and L. Gold, *Systematic evolution of ligands by exponential enrichment: RNA ligands to bacteriophage T4 DNA polymerase.* Science, 1990. **249**(4968): p. 505-510.
26. Schneider, T.D., et al., *Information content of binding sites on nucleotide sequences.* J Mol Biol, 1986. **188**(3): p. 415-31.
27. Harbison, C.T., et al., *Transcriptional regulatory code of a eukaryotic genome.* Nature, 2004. **431**(7004): p. 99-104.
28. Gama-Castro, S., et al., *RegulonDB (version 6.0): gene regulation model of Escherichia coli K-12 beyond transcription, active (experimental) annotated promoters and Textpresso navigation.* Nucleic acids research, 2008. **36**(suppl 1): p. D120-D124.
29. Weirauch, M.T., et al., *Determination and inference of eukaryotic transcription factor sequence specificity.* Cell, 2014. **158**(6): p. 1431-1443.
30. Jolma, A., et al., *DNA-binding specificities of human transcription factors.* Cell, 2013. **152**(1): p. 327-339.
31. Grove, C.A., et al., *A multiparameter network reveals extensive divergence between C. elegans bHLH transcription factors.* Cell, 2009. **138**(2): p. 314-327.
32. Badis, G., et al., *Diversity and complexity in DNA recognition by transcription factors.* Science, 2009. **324**(5935): p. 1720-1723.
33. Noyes, M.B., et al., *Analysis of homeodomain specificities allows the family-wide prediction of preferred recognition sites.* Cell, 2008. **133**(7): p. 1277-1289.
34. Berger, M.F., et al., *Variation in homeodomain DNA binding revealed by high-resolution analysis of sequence preferences.* Cell, 2008. **133**(7): p. 1266-1276.
35. Mathelier, A., et al., *JASPAR 2014: an extensively expanded and updated open-access database of transcription factor binding profiles.* Nucleic acids research, 2013: p. gkt997.
36. Cipriano, M.J., et al., *RegTransBase–a database of regulatory sequences and interactions based on literature: a resource for investigating transcriptional regulation in prokaryotes.* BMC genomics, 2013. **14**(1): p. 213.
37. Savir, Y., et al., *Balancing speed and accuracy of polyclonal T cell activation: a role for extracellular feedback.* BMC Syst Biol, 2012. **6**: p. 111.
38. Savir, Y. and T. Tlusty, *The Ribosome as an Optimal Decoder: A Lesson in Molecular Recognition.* Cell, 2013. **153**(2): p. 471-479.
39. Savir, Y. and T. Tlusty, *RecA-mediated homology search as a nearly optimal signal detection system.* Mol Cell, 2010. **40**(3): p. 388-96.
40. Savir, Y., et al., *Cross-species analysis traces adaptation of Rubisco toward optimality in a low-dimensional landscape.* Proc Natl Acad Sci U S A, 2010. **107**(8): p. 3475-80.

41. Salgado, H., et al., *RegulonDB v8. 0: omics data sets, evolutionary conservation, regulatory phrases, cross-validated gold standards and more.* Nucleic acids research, 2013. **41**(D1): p. D203-D213.
42. Zhu, C., et al., *High-resolution DNA-binding specificity analysis of yeast transcription factors.* Genome research, 2009. **19**(4): p. 556-566.

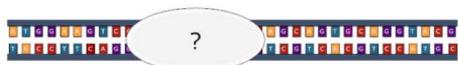
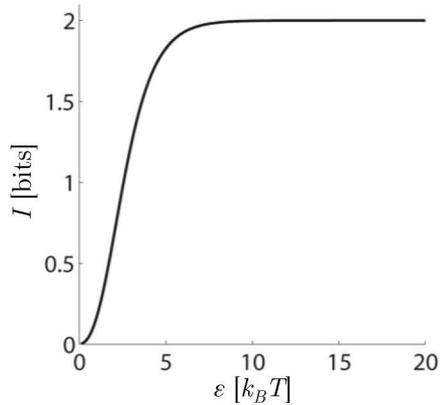
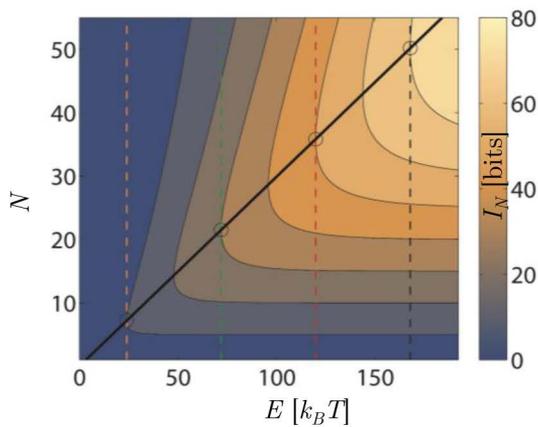
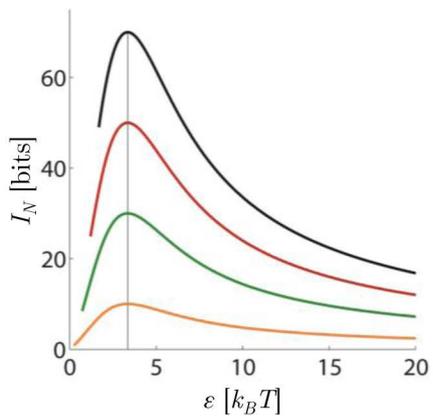
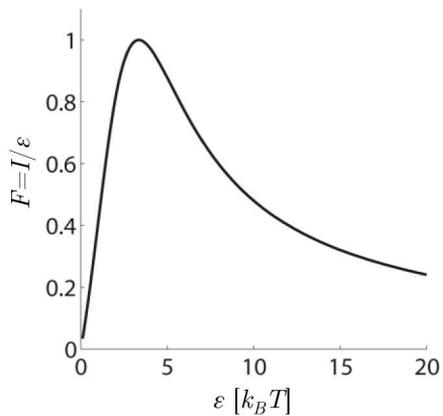

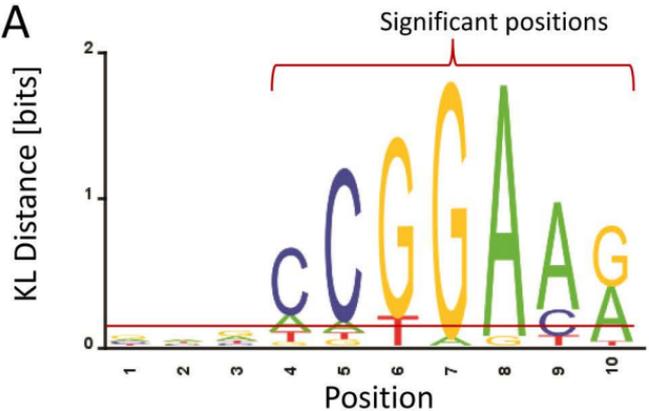 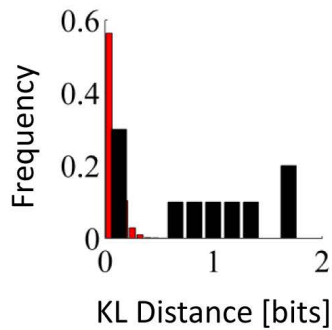 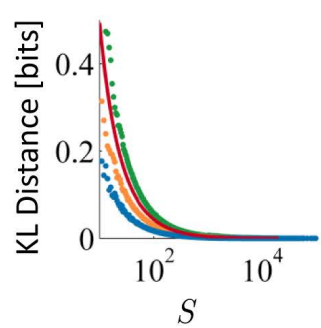 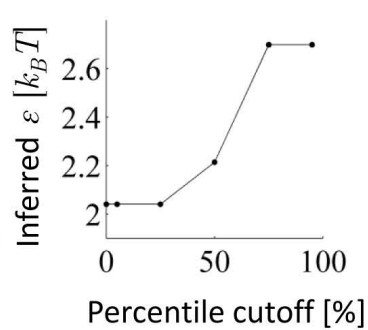

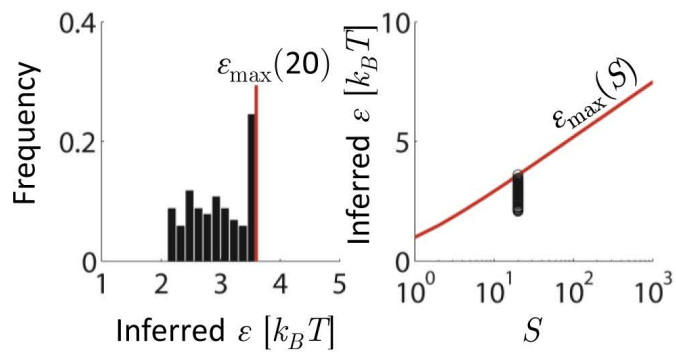 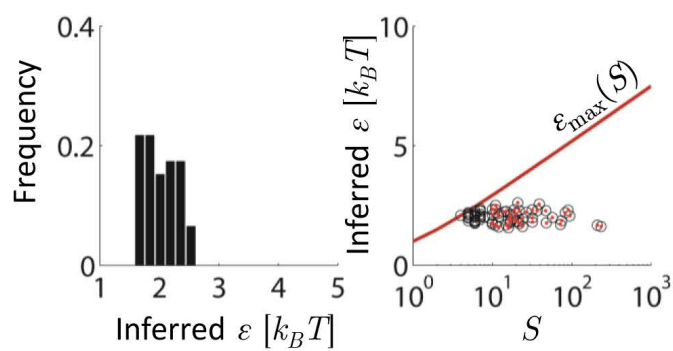

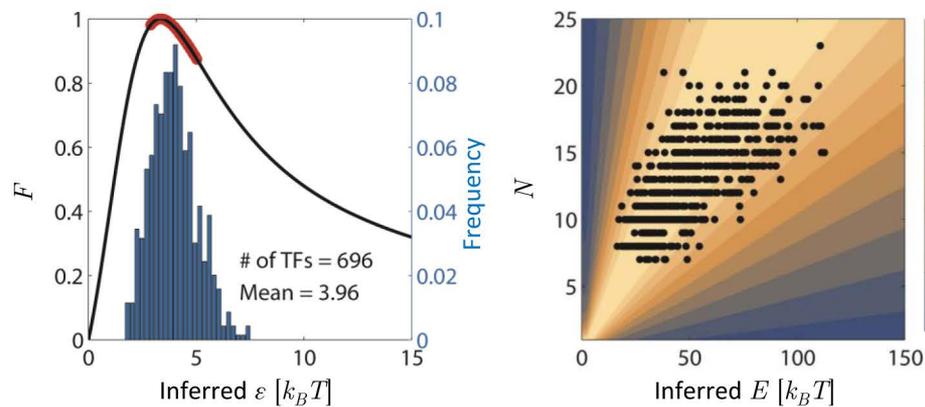
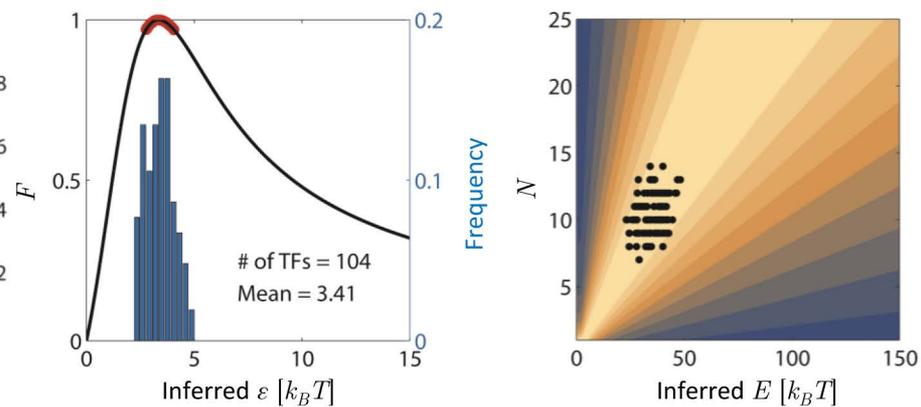
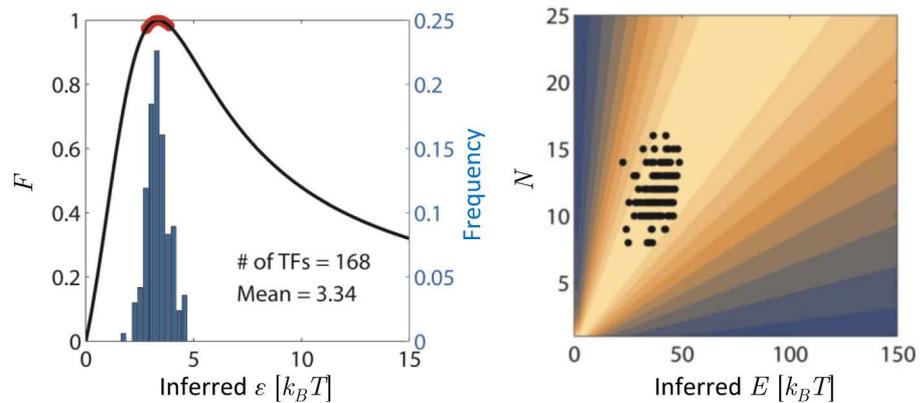
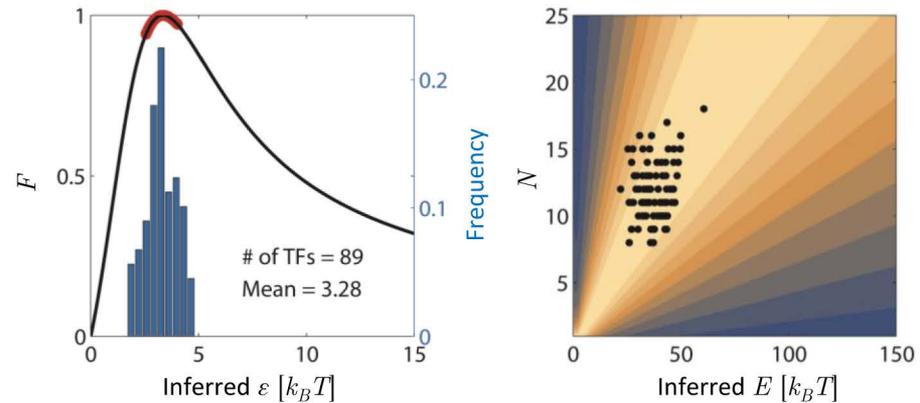

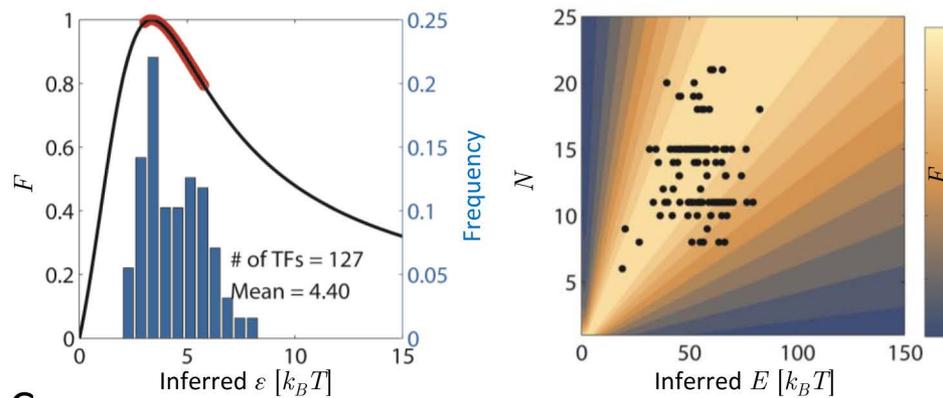

A. Jaspar vertebrates

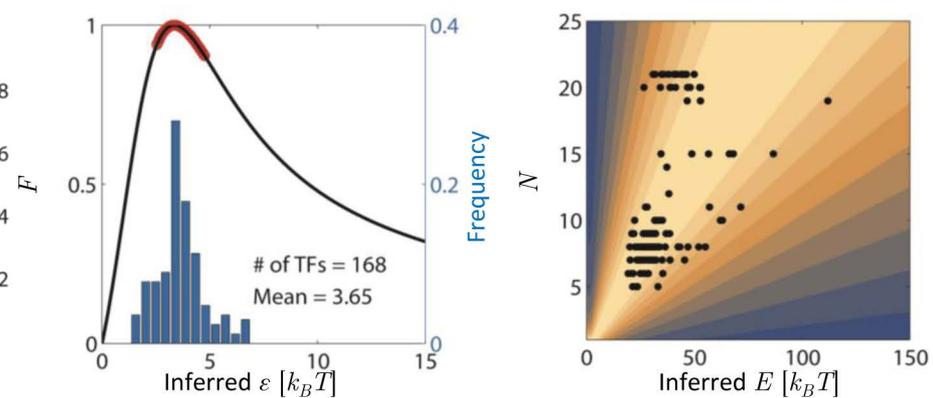

B. Jaspar fungi

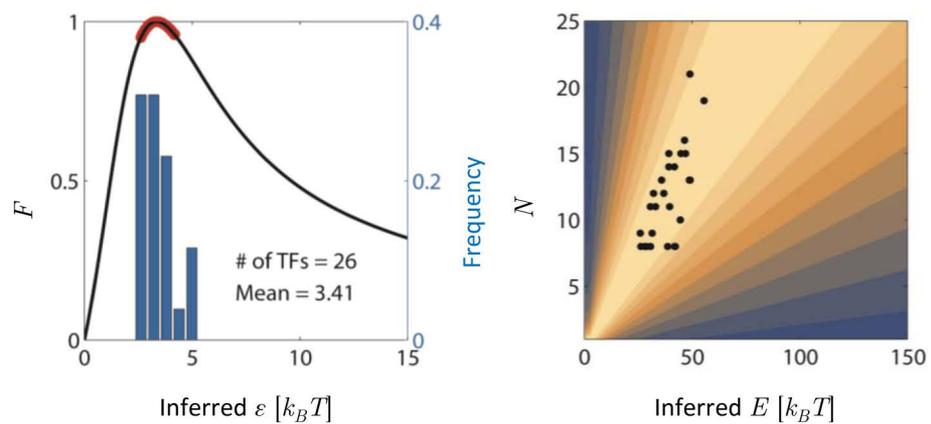

C. Jaspar plants

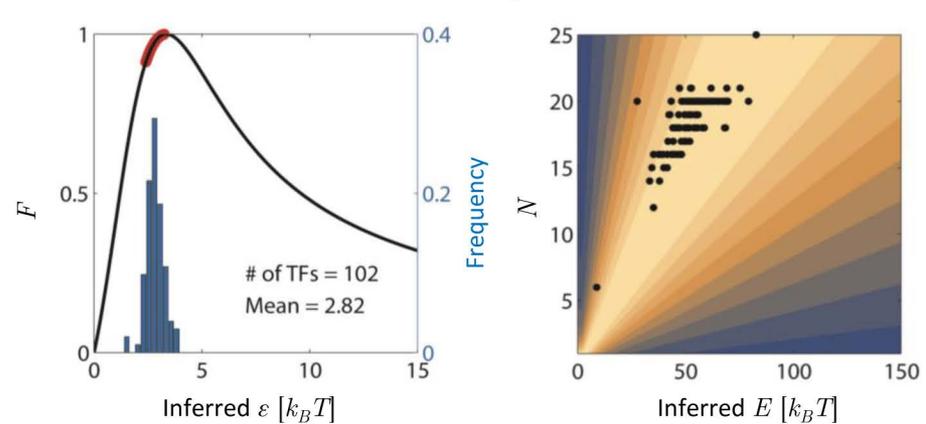

D. Bacteria, RegTransBase

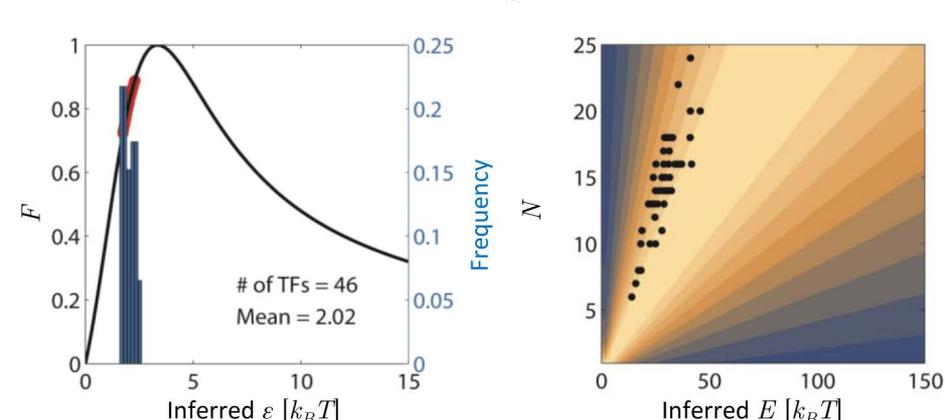

E. Bacteria RegulonDB

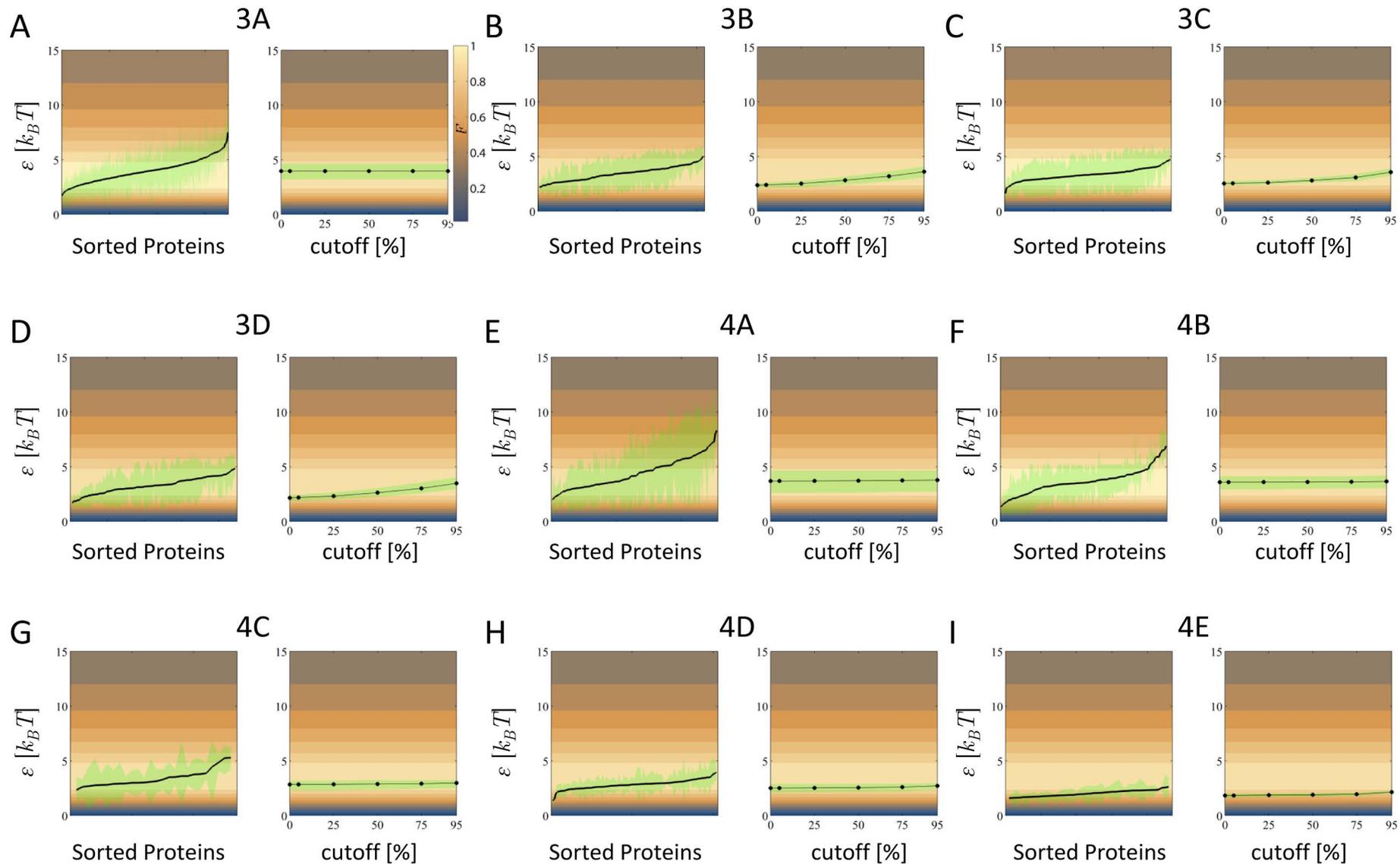